\documentclass[
   prd,
   amsfonts,
   amssymb,
   amsmath,
   superscriptaddress,
   twocolumn,
   eqsecnum,
   showpacs,
   preprintnumbers
   ]{revtex4}
\usepackage{color}
\usepackage{graphicx}
\usepackage{subfigure}
\usepackage{time}
\usepackage[pdftex,colorlinks=true]{hyperref}
\newcommand{\oneplusone}{(1+1)}            
\newcommand{\threeplusone}{(3+1)}          

\newcommand{\etal}{\textit{et~al.}}        
\newcommand{\rd}{\mathrm{d}}               
\newcommand{\rD}{\mathrm{D}}               
\newcommand{\defby}{\equiv}                
\newcommand{\pk}{k}                        
\newcommand{\kperp}{k_{\perp}}             
\newcommand{\keta}{k_{\eta}}               
\newcommand{\calL}{\mathcal{L}}            
\newcommand{\calE}{\mathcal{E}}            
\newcommand{\calP}{\mathcal{P}}            
\newcommand{\Diag}[1]{\mathrm{diag}( \, #1 \, )}   
\newcommand{\Set}[1]{( \, #1 \, )}                 
\newcommand{\Expect}[1]
   {\ensuremath{\langle \, #1 \,  \rangle}}
\newcommand{\Comm}[2]
   {\ensuremath{[ \, #1, #2 \, ]}}
\newcommand{\AntiComm}[2]
   {\ensuremath{\{ \, #1, #2 \, \}}}
\newcommand{\PB}[2]
   {\ensuremath{\{ \, #1, #2 \, \}}}
\begin{document}
%
%
%
\preprint{LA-UR-09-03554}
\title[Boltzmann-Vlasov solutions]
      {Fermion particle production in semi-classical
       Boltzmann-Vlasov transport theory}

\author{John F. Dawson}
\email{john.dawson@unh.edu}
\affiliation{Department of Physics,
   University of New Hampshire,
   Durham, NH 03824}

\author{Bogdan Mihaila}
\email{bmihaila@lanl.gov}
\affiliation{Materials Science and Technology Division,
   Los Alamos National Laboratory,
   Los Alamos, NM 87545}

\author{Fred Cooper}
\email{cooper@santafe.edu}
\affiliation{National Science Foundation,
   4201 Wilson Blvd.,
   Arlington, VA 22230}
\affiliation{Santa Fe Institute,
   Santa Fe, NM 87501}
\affiliation{Center for Nonlinear Studies,
   Los Alamos National Laboratory,
   Los Alamos, NM 87545}



\begin{abstract}
We present numerical solutions of the semi-classical Boltzmann-Vlasov equation for fermion particle-antiparticle production by strong electric fields in boost-invariant coordinates in \oneplusone\ and \threeplusone\ dimensional QED.  We compare the Boltzmann-Vlasov results with those of recent quantum field theory calculations and find good agreement. We conclude that extending the Boltzmann-Vlasov approach to the case of QCD should allow us to do a thorough investigation of how back-reaction affects recent results on the dependence of the transverse momentum distribution of quarks and anti-quarks on a second Casimir invariant of color SU(3).  \end{abstract}
\maketitle
%
%
\section{Introduction}
\label{s:intro}

In recent papers, we have presented numerical quantum field theory calculations of the dynamics of fermion pair production by strong electric fields with back-reaction in boost-invariant coordinates in \oneplusone\ and \threeplusone\ dimensions \cite{r:Mihaila:2008dp,r:Mihaila:2009tg}.  The purpose of the present paper is to compare these calculations with the results of numerical calculations using a semi-classical Boltzmann-Vlasov (BV) equation  with a Schwinger source term for particle pair creation.  We find that in \threeplusone\ dimensions this semi-classical transport approximation works even better than it did in \oneplusone\ dimensions.  With the confidence that this model is working well for \threeplusone\ dimensional quantum electrodynamics (QED), our program is to extend this calculation to quantum chromodynamics (QCD), where recently it has been shown that the WKB source term used by previous studies of pair production using the BV equation neglected an important term which depends on the second Casimir invariant of SU(3)~\cite{ref:NayakNei,r:Nayak:2005uq}. The BV equation is much quicker to implement than the full field theory calculation and will let us explore the parameter space quickly before we perform more computer-intensive field theory calculations.

The model we are using for the production of the particles following a heavy ion collision is the so-called color flux tube model. The color flux tube model assumes that when two relativistic  heavy ions collide multiple gluons are exchanged which leads to the formation of a strong color electric field.  This model was studied extensively in the 1980's by several authors. These include Bialas, \etal\ \cite{r:Bialas:1984wd,r:Bialas:1985nx,r:Bialas:1985eu,r:Bialas:1985oq,r:Bialas:1986qe,r:Bialas:1988kl} and by Kajantie and Matsui \cite{r:Kajantie:1985ai}. The idea of using a boost invariant Bolzmann-Vlasov equation to study the time evolution of the plasma formed by the produced quarks and gluons was first put forward by Bialis and Czyz~\cite{r:Bialas:1984wd} and this was then generalized to include a Schwinger source term by Gatoff, Kerman and Matsui~\cite{r:Gatoff:1987fk}. At that time the validity of the BV approach was not known.  However, once field theory calculations of this process were done in the 1990's~\cite{r:Cooper:1993uq}, it was clear that solving the BV equations with a Schwinger source term was a reasonable approximation.  In the original work on QCD, the source term used was a WKB source term proposed by Casher, Neuberger and Nussinov~\cite{r:Casher:1979fk}, which recently has been shown to be incorrect by Nayak and collaborators~\cite{ref:NayakNei,r:Nayak:2005uq}.  For constant chromoelectric fields the dependence on the second Casimir invariant can affect the transverse distribution of produced particles by as much as 15\%~\cite{Cooper:2008tg} which is a reason to correctly formulate the transport approach for the QCD plasma evolution and compare it to the field theory calculation.

Our discussion of the BV equation in boost invariant coordinates for \threeplusone\ dimensional QED  follows closely in spirit  work by Kluger, \etal\ \cite{r:KESCMprl91,r:Kluger:1992fk} and by Cooper, \etal\ \cite{r:Cooper:1993uq}. We  follow the method of solution used in these previous papers.
In Section~\ref{s:classtheory}, we discuss the classical theory for the boost-invariant coordinate system which we use in this paper and develop the equations needed for solutions of the BV equation.  Numerical methods and results are discussed in Section~\ref{s:numerical}, and conclusions given in Section~\ref{s:conclusions}.

%
%
\section{Classical theory}
\label{s:classtheory}

We wish to describe the dynamics of a relativistic particle of mass $M$ and charge $e$ interacting with an electromagnetic field in an arbitrary coordinate system.  Let $x^{\mu}(s)$ be the trajectory of a particle in space-time described parametrically by the arc-length $\rd s$, defined by
\begin{equation}\label{e:dsdef}
   ( \rd s )^2
   =
   g_{\mu,\nu}(x) \, \rd x^{\mu} \rd x^{\mu} \>.
\end{equation}
The velocity four-vector $u^{\mu}(s)$ along the trajectory curve is given by
\begin{equation}\label{e:udef}
   u^{\mu}(s)
   \defby
   \frac{ \rd x^{\mu}(s) }{ \rd s } \>,
   \quad
   u^{\mu}(s) u_{\mu}(s)
   =
   1 \>,
\end{equation}
and the Lagrangian is
\begin{equation}\label{e:Lagrangiandef}
   \calL[ \, x^\mu, u^{\mu} \, ]
   =
   \frac{1}{2} \, M \, u^{\mu}(s) u_{\mu}(s)
   +
   e \, u^{\mu}(s) A_{\mu}(x) \>.
\end{equation}
The canonical momenta $p_{\mu}(s)$ is given by
\begin{equation}\label{e:pcandef}
   p_{\mu}(s)
   \defby
   \frac{\partial \calL}{\partial u^{\mu}}
   =
  k_{\mu}(s)
   +
   e \, A_{\mu}(x) \>,
\end{equation}
where $k_{\mu}(s) = M \, u_{\mu}(s)$ is the kinetic momentum.  In terms of the kinetic momentum, Lagrange's equation give
\begin{equation}\label{e:eomI}
   M \, \frac{\rd \pk_{\mu}(s)}{\rd s}
   =
   e \, F_{\mu\nu}(x) \, \pk^{\nu}(s) \>,
\end{equation}
where $F_{\mu\nu}(x) = \partial_{\mu} A_{\nu}(x) - \partial_{\nu} A_{\mu}(x)$ is the field tensor, which satisfies the Maxwell equations,
\begin{equation}\label{e:Maxwell}
   \frac{1}{\sqrt{-g}} \,
   \partial_{\mu}
   \bigl [ \,
      \sqrt{-g} \, F^{\mu\nu}(x) \,
   \bigr ]
   =
   J^{\nu}(x) \>,
\end{equation}
where the current is the sum of convective and polarization currents.  The classical convective current is given by
\begin{equation}\label{e:conJ}
   J^{\text{con}\,\mu}(x)
   =
   e
   \sum \int \rd s \, u^{\mu}(s) \, \delta^4[ \, x - x(s) \, ] \>,
\end{equation}
where $x(s)$ is a solution of the equations of motion, and the sum goes over all species, particles, antiparticles, and spins.  The energy momentum tensor densities for the particles $t^{\mu\nu}(x)$ and field $\Theta^{\mu\nu}(x)$ are given by
\begin{subequations}\label{e:enrgmompartfield}
\begin{align}
   t^{\mu\nu}(x)
   &=
   \sum \int \rd s \, u^{\mu}(s) \, k^{\nu}(s) \,
   \delta^4[ \, x - x(s) \, ] \>,
   \label{e:enrgmompart} \\
   \Theta^{\mu\nu}(x)
   &=
   \frac{1}{4} \,
   g_{\mu\nu} \, F^{\alpha\beta} F_{\alpha\beta}
   +
   F_{\mu\alpha} g^{\alpha\beta} F_{\beta\nu} \>.
   \label{e:enrgmomfield}
\end{align}
\end{subequations}
The field energy-momentum tensor density satisfies
\begin{equation}\label{e:fieldenergymomcons}
   \frac{1}{\sqrt{-g}} \,
   \partial_{\mu}
   \bigl [ \,
      \sqrt{-g} \, \Theta^{\mu\nu}(x) \,
   \bigr ]
   =
   -
   F^{\nu\sigma}(x) \, J_{\sigma}(x) \>.
\end{equation}

%
%
\subsection{Trajectory solutions}
\label{ss:trajectories}

We next find trajectory solutions to the equations of motion in boost invariant coordinates.  The Cartesian set of coordinates is designated by Roman letters: $x^{a} = \Set{t,x,y,z}$, with the metric $\eta_{ab} = \Diag{1,-1,-1,-1}$.  Boost-invariant variables are designated by Greek letters: $x^{\mu} = \Set{\tau,\rho,\theta,\eta}$, where
\begin{alignat}{2}
   t & = \tau \cosh \eta \>, & \qquad z & = \tau \sinh \eta \>,
   \label{e:xdefs} \\
   x & = \rho \cos\theta \>, & \qquad y & = \rho \sin\theta \>,
   \notag
\end{alignat}
with the metric
\begin{equation*}
   g_{\mu\nu}(x)
   =
   \Diag{ 1, -1, -\rho^2, - \tau^2 } \>.
\end{equation*}
The kinetic momentum in boost-invariant coordinates is then
given by
\begin{align}
   \pk^{\mu}
   &=
   \Set{ \pk^{\tau}, \pk^{\rho}, \pk^{\theta}, \pk^{\eta} }
   \label{e:kcontadef} \\
   &=
   M \, \frac{ \rd x^{\mu}(s) }{ \rd s }
   =
   M \,
   \Set{ \tau', \rho', \theta', \eta' } \>.
   \notag
\end{align}
Here a primed quantity means a derivative with respect to~$s$.  The mass shell restriction requires
\begin{equation}\label{e:klength}
   \pk^{\mu} \pk_{\mu}
   =
   \pk_{\tau}^2
   -
   \kperp^2
   -
   [ \, \pk_{\eta}/\tau \, ]^{2}
   =
   M^2 \>,
\end{equation}
where we have defined $\kperp$ by
\begin{equation}\label{e:pkperpdef}
   \kperp^2
   =
   \pk_{\rho}^{2}
   +
   [ \, \pk_{\theta}/\rho \, ]^{2}
   =
   M^2 \,
   [ \, \rho'^2 + \rho^2 \theta'^2 \, ] \>.
   \notag
\end{equation}
So $k_{\tau} = \omega_{\pk_{\perp},\pk_{\eta}}(\tau)$, where
\begin{equation}\label{e:omegadef}
   \omega_{\pk_{\perp},\pk_{\eta}}(\tau)
   =
   \sqrt{ \kperp^2 + [ \, \pk_{\eta}/\tau \, ]^{2} + M^2 } \>.
\end{equation}

We restrict the vector potential and electric fields to be in the $\eta$-direction and depend only on $\tau$, i.e.: $A_{\mu}(x) = \Set{0,0,0,A_{\eta}(\tau)}$.  Then the only non-vanishing components of the field tensor are given by
\begin{equation}\label{e:Fcontradef}
   F_{\tau,\eta}(x)
   =
   - F_{\eta,\tau}(x)
   =
   \partial_{\tau} A_{\eta}(\tau)
   =
   -
   \tau E(\tau) \>.
\end{equation}
Here we have defined $E(\tau) = - [ \partial_{\tau} A_{\eta}(\tau) ]/ \tau$.  So then the Newton's Eqs.~\eqref{e:eomI} become
\begin{subequations}\label{e:eomII}
\begin{align}
   M \, \frac{\rd k_{\tau}(s)}{\rd s}
   &=
   e \, E(\tau) \, k_{\eta}(s) / \tau \>,
   \label{e:eomIIa} \\
   M \, \frac{\rd k_{\eta}(s)}{\rd s}
   &=
   e \, \tau \, E(\tau) \, k_{\tau}(s) \>,
   \label{e:eomIIb}
\end{align}
\end{subequations}
with $k_{\rho}$ and $k_{\theta}$ constants of the motion.  Using the fact that $k_{\tau}(s) = M \, \rd \tau/ \rd s$, Eq.~\eqref{e:eomIIb} becomes
\begin{equation}\label{e:eomIIc}
   \frac{\rd}{\rd \tau} \,
   [ \, k_\eta(\tau) + e \, A_\eta(\tau) \, ]
   =
   0 \>,
\end{equation}
from which we conclude that
$p_\eta= k_\eta(\tau) + e \, A_\eta(\tau)$ is a constant of the motion.  We can also define $x$- and $y$-components of the transverse momentum by
\begin{align}
   \pk_{x}
   &\defby
   \pk^{\rho} \cos \theta
   -
   \rho \, \pk^{\theta} \sin \theta
   \label{e:pkxdef} \\
   &=
   M \,
   \bigl [ \,
      \rho' \, \cos \theta
      -
      \rho \, \theta' \, \sin \theta \,
   \bigr ]
   \defby
   M \, x' \>,
   \notag \\
   \pk_{y}
   &\defby
   \pk^{\rho} \sin \theta
   +
   \rho \, \pk^{\theta} \cos \theta
   \label{e:pkydef} \\
   &=
   M \,
   \bigl [ \,
      \rho' \, \sin \theta
      +
      \rho \, \theta' \, \cos \theta \,
   \bigr ]
   \defby
   M \, y' \>.
   \notag
\end{align}
In cylindrical coordinates,
\begin{equation}\label{e:kxkykperpphi}
   \pk_{x}
   =
   \kperp \cos \phi \>,
   \qquad
   \pk_{y}
   =
   \kperp \sin \phi \>,
\end{equation}
which defines the angle $\phi$.  By computing the Jacobians of these transformations, we show that volume elements are related by
\begin{equation}\label{e:pkvolelements}
   \rd \pk_x \rd \pk_y
   =
   \kperp \rd \kperp \, \rd \phi
   =
   \frac{ \rd \pk_{\rho} \, \rd \pk_{\theta} }{ \rho } \>.
\end{equation}

%
%
\subsection{Rapidity variables}
\label{ss:rapidity}

It will be useful to define rapidity momentum variables $(r,y)$.  These variables are defined by
\begin{equation}\label{e:rapiditydefs}
   k_{t}
   =
   r \cosh y \>,
   \qquad
   k_{z}
   =
   r \sinh y \>,
\end{equation}
which can be related to our boost-invariant set $( k_{\tau}, k_{\eta} )$ by
\begin{equation}\label{e:kmutoyM}
   k_{\mu}
   =
   M \frac{\rd x_{\mu}}{\rd s}
   =
   \frac{\partial x_{\mu}}{\partial x_{a}} \, k_{a} \>,
\end{equation}
from which we find
\begin{subequations}\label{e:ktauketatorapidity}
\begin{align}
   k_{\tau}
   &=
   r \cosh( \eta - y ) \>,
   \label{e:ktaurapid} \\
   k_{\eta} / \tau
   &=
   r \sinh( \eta - y ) \>.
   \label{e:ketarapid}
\end{align}
\end{subequations}
On the energy shell, we have
\begin{equation}\label{e:energyshell}
   k_{\tau}^2
   -
   k_{\perp}^2
   -
   [ \, k_{\eta} / \tau ]^2
   =
   r^2 \,
   -
   k_{\perp}^2
   =
   M^2 \>,
\end{equation}
so on the energy shell, $r = M_{\perp} \equiv \sqrt{k_{\perp}^2 + M^2}$.  The Jacobian for this transformation is given by
\begin{equation}\label{e:Jacobian}
   \begin{vmatrix}
      \partial k_{\tau} / \partial r \>,
      &
      \partial k_{\tau} / \partial y
      \\
      \partial k_{\eta} / \partial r \>,
      &
      \partial k_{\eta} / \partial y
   \end{vmatrix}
   =
   \tau r \>,
\end{equation}
so
\begin{equation}\label{e:dktaudketatorapidity}
   \rd k_{\tau} \, ( \, \rd k_{\eta} / \tau \, )
   =
   r \, \rd r \, \rd y \>.
\end{equation}
We will use this result in Section~\ref{ss:BoltzmannVlasov} below.

%
%
\subsection{The Boltzmann-Vlasov equation}
\label{ss:BoltzmannVlasov}

We define a particle distribution function $f(x,k)$ such that the particle current density is given by (see, for example, Calzetta and Hu \cite{r:Calzetta:2008pb})
\begin{equation}\label{e:pcurrentdef}
   N^{\mu}(x)
   =
   \int \rD k \, k^{\mu} \, f(x,k) \>,
\end{equation}
and the particle energy-momentum density tensor is given by
\begin{equation}\label{e:penergymom}
   t^{\mu\nu}(x)
   =
   \int \rD k \, k^{\mu} k^{\nu} \, f(x,k) \>,
\end{equation}
where
\begin{equation}\label{e:Dkdef}
   \rD k
   =
   \frac{2R \, \Theta(k_0) \, \delta( k^2 - M^2 ) \, \rd^4 k}
        {(2\pi)^3 \, \sqrt{-g}} \>,
\end{equation}
with $R$ a degeneracy factor.  For a single species of fermions in \threeplusone dimensions, counting particles, antiparticles, and spin, $R=4$.  In a general coordinate system, the BV equation is given by (see for example Cooper, \etal\ \cite{r:Cooper:1993uq} or Gatoff, \etal\ \cite{r:Gatoff:1987fk}.)
\begin{equation}\label{e:BVeqI}
   k^{\mu} \,
   \Bigl \{ \,
      \frac{\partial}{\partial x^{\mu}}
      -
      e \, F_{\mu\nu}(x) \, \frac{\partial}{\partial k_{\nu}} \,
   \Bigr \} \, f(x,k)
   =
   k^{0} \, C(x,k) \>,
\end{equation}
where $C(x,k)$ is a source term.  Multiplying \eqref{e:BVeqI} by $\sqrt{-g}$ and integrating over $\rD k$ gives
\begin{equation}\label{e:Ncons}
   \frac{1}{\sqrt{-g}} \,
   \partial_{\mu}
   \bigl [ \,
      \sqrt{-g} \, N^{\mu}(x) \,
   \bigr ]
   =
   C(x) \>,
\end{equation}
where
\begin{equation}\label{e:Cxdef}
   C(x)
   =
   \int \, \rD k \, k^{0} \, C(x,k) \>.
\end{equation}
So if $C(x,k) = 0$, particle number is conserved.  Multiplying \eqref{e:BVeqI} by $k^{\nu} \, \sqrt{-g}$ and integrating over $\rD k$ gives
\begin{equation}\label{e:tconscov}
   \frac{1}{\sqrt{-g}} \,
   \partial_{\mu}
   \bigl [ \,
      \sqrt{-g} \, t^{\mu\nu}(x) \,
   \bigr ]
   -
   F^{\nu\sigma}(x) \, J_{\sigma}(x)
   =
   C^{\nu}(x) \>,
\end{equation}
where
\begin{equation}\label{e:Cmuxdef}
   C^{\nu}(x)
   =
   \int \, \rD k \, k^{0} k^{\nu} \, C(x,k) \>,
\end{equation}
and $J_{\sigma}(x) = e N_{\sigma}(x)$.  So if $C(x,k)=0$, combining Eqs.~\eqref{e:fieldenergymomcons} and \eqref{e:tconscov}, we see that with no source term, the total energy-momentum tensor density,
\begin{equation}\label{e:totenergymomdef}
   T^{\mu\nu}(x)
   =
   t^{\mu\nu}(x) + \Theta^{\mu\nu}(x) \>,
\end{equation}
satisfies a conservation law, $T^{\mu\nu}{}_{;\mu}(x)=0$.  For our case, the source of particles is creation of particle-hole pairs via the Schwinger mechanism, so the particle number is \emph{not} conserved and the energy-momentum tensor, using only convective currents, is also not conserved.

In boost-invariant coordinates, we assume that the distribution function is a function of $f(\tau,\kperp,\keta)$ only.  So choosing a surface element in the direction of constant $\tau$, we have $\rd \Sigma = \tau \, \rd^2 x_{\perp} \rd \eta$ where $\rd^2 x_{\perp} = \rho \, \rd \rho \, \rd \theta$ is the perpendicular area, and
\begin{equation}\label{e:Nbi}
   N^{\mu}(\tau)
   =
   \frac{R}{(2\pi)^3} \,
   \iint \! \rd^2 \kperp
   \int_{-\infty}^{+\infty} \!\!\!\! \rd \keta \,
   \frac{k^{\mu} \, f(\tau,\kperp,\keta)}
        {\tau \, \omega_{\kperp,\keta}(\tau)} \>,
\end{equation}
where $\rd^2 \kperp = \kperp \rd \kperp \, \rd \phi$ and $\omega_{\kperp,\keta}(\tau)$ is given by Eq.~\eqref{e:omegadef}.  The $\mu=0$ component of \eqref{e:Nbi} gives the number of particles per unit ``volume'' in boost-invariant coordinates:
\begin{align}
   \frac{\rd^3 N(\tau)}{\rd^2 x_{\perp} \rd \eta}
   &=
   \tau \, N_{0}(\tau)
   \label{e:Ndefbi} \\
   &=
   \frac{R}{(2\pi)^3} \,
   \iint \! \rd^2 \kperp
   \int_{-\infty}^{+\infty} \!\!\!\! \rd \keta \,
   f(\tau,\kperp,\keta) \>,
   \notag
\end{align}
In terms of rapidity variables, Eq.~\eqref{e:Ndefbi} becomes
\begin{align}
   \rd^6 N
   &=
   \frac{R}{(2\pi)^3} \,
   [ \, \tau \, \rd^2 x_{\perp} \rd \eta \, ] \,
   [ \, \rd^2 k_{\perp} \rd y \, ]
   \label{e:dNsixR}
   \\ &\qquad \times
   \int \frac{ r \, \rd r \, 2 \Theta(r) }
             { 2 r } \,
   \delta( r - M_{\perp} ) \, k_{\tau} \,
   f(\tau,k_{\perp},k_{\eta})
   \notag \\
   &=
   \frac{R}{(2\pi)^3} \,
   [ \, \tau \, \rd^2 x_{\perp} \rd \eta \, ] \,
   [ \, \rd^2 k_{\perp} \rd y \, ] \,
   \omega_{k_{\perp},k_{\eta}} \,
   f(\tau,k_{\perp},k_{\eta}) \>.
   \notag
\end{align}
So the momentum distribution in rapidity variables is given by
\begin{align}
   &\frac{\rd^5 N}{\rd^2 x_{\perp} \rd^2 k_{\perp} \rd y}
   =
   \frac{R \tau}{(2\pi)^3} \,
   \int_{-\infty}^{+\infty} \!\!\!\! \rd \eta \,
      \omega_{k_{\perp},k_{\eta}} \,
   f(\tau,k_{\perp},k_{\eta})
   \label{e:d5Nrapid} \\
   & \quad =
   \frac{R \tau}{(2\pi)^3} \,
   \int_{-\infty}^{+\infty} \!\!\!\! \rd k_{\eta} \,
   \Bigl | \, \frac{\partial \eta}{\partial k_{\eta}} \, \Bigr | \,
      \omega_{k_{\perp},k_{\eta}} \,
   f(\tau,k_{\perp},k_{\eta}) \>.
   \notag
\end{align}
But the only $\eta$ dependence is through $k_{\eta}$.  Evaluating Eqs.~\eqref{e:ktauketatorapidity} on the mass shell, we find
\begin{equation}\label{e:massshelleqs}
   \omega_{k_{\perp},k_{\eta}}
   =
   M_{\perp} \, \cosh( \eta - y ) \>,
   \quad
   k_{\eta}
   =
   \tau \, M_{\perp} \, \sinh( \eta - y ) \>,
\end{equation}
so that for fixed $y$ and $k_{\perp}$, we find
\begin{equation}\label{e:jacobrap}
   \frac{\partial k_{\eta}}{\partial \eta}
   =
   \tau \, \omega_{k_{\perp},k_{\eta}} \>,
\end{equation}
and \eqref{e:d5Nrapid} becomes
\begin{equation}\label{e:dN5rapidII}
   \frac{\rd^5 N}{\rd^2 x_{\perp} \rd^2 k_{\perp} \rd y}
   =
   \frac{R}{(2\pi)^3} \,
   \int_{-\infty}^{+\infty} \!\!\!\! \rd k_{\eta} \,
   f(\tau,k_{\perp},k_{\eta}) \>,
\end{equation}
and is independent of rapidity.

\begin{figure}[t!]
   \centering
   \includegraphics[width=\columnwidth]{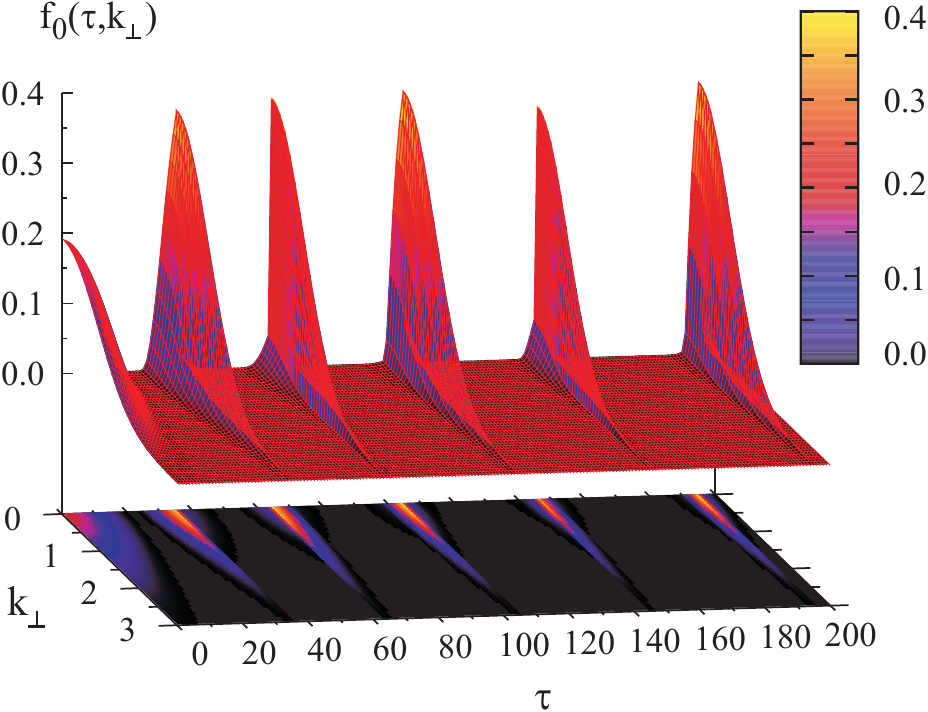}
   \caption{\label{f:0}(Color online)
   Evolution of the distribution function $f(\tau,\kperp,0)$ for a
   typical case with $M=1$, $e=1$, $A(\tau_0)=0$ and $E(\tau_0)=4$.}
\end{figure}

Adding a Schwinger source term to the BV equation, in boost-invariant coordinates the only non-vanishing components of $F_{\mu\nu}(x)$ in our case are given in Eq.~\eqref{e:Fcontradef}, so that Eq.~\eqref{e:BVeqI} becomes
\begin{equation}\label{e:BVeqII}
   \Bigl \{ \,
      \frac{\partial}{\partial \tau}
      -
      e \frac{\partial A(\tau)}{\partial \tau}  \,
      \frac{\partial}{\partial \keta} \,
   \Bigr \} \,  f(\tau,\kperp,\keta)
   =
   C(\tau,\kperp,\keta) \>,
\end{equation}
where the source term is given by
\begin{equation}\label{e:Cbi}
   C(\tau,\kperp,\keta)
   =
   P(\tau,\kperp) \,
   | e E(\tau) | \,
   S(\tau,\kperp) \,
   \delta( \keta/\tau ) \>,
\end{equation}
with $P(\tau,\kperp)$ a Pauli suppression factor evaluated at
$\keta = 0$,
\begin{equation}\label{e:Pauli}
   P(\tau,\kperp)
   =
   1 - 2 \, f(\tau,\kperp,0) \>,
\end{equation}
and $S(\tau,\kperp)$ is the Schwinger pair creation factor
\begin{equation}\label{e:schwingerII}
   S(\tau,\kperp)
   =
   - \ln
   \Bigl [ \,
      1
      -
      e^{ - \pi ( \kperp^2 + M^2 ) / |e E(\tau) | } \,
   \Bigr ]  \>.
\end{equation}
We solve Eq.~\eqref{e:BVeqII} for $f(\tau,\kperp,\keta)$ using the method of characteristics.  In Section~\ref{ss:trajectories}, we found the particle trajectories and we showed that $\keta(\tau) = p_{\eta} - e \, A_{\eta}(\tau)$, where $p_\eta$ is a constant of the motion.  So the total derivative of $f[\tau,\kperp,\keta(\tau)]$ with respect to $\tau$ is given by
\begin{align*}
   &\frac{ \rd f[ \tau,\kperp,\keta(\tau) ] }
        { \rd \tau }
   \\
   & \qquad
   =
   \frac{\partial f[\tau,\kperp,\keta(\tau)]}{\partial \tau}
   -
   e \frac{\partial A_{\eta}(\tau)}{\partial \tau} \,
   \frac{\partial f[\tau,\kperp,\keta(\tau)]}
        {\partial \keta} \>.
   \notag
\end{align*}
Assuming that $f(\tau_0,\kperp,\keta) = 0$, we then have
\begin{align}
   &f( \tau,\kperp,\keta )
   =
   \int_{\tau_0}^{\tau} \rd \tau' \,
   \tau' \,
   P(\tau',\kperp) \,
   | e E(\tau') | \,
   S(\tau',\kperp)
   \notag \\ & \qquad\qquad \times
   \delta[ \keta + e A_{\eta}(\tau) - e A_{\eta}(\tau') ] \>,
   \label{e:fsol}
\end{align}
which can be integrated to give
\begin{align}
   f(\tau,\kperp,\keta)
   &=
   \sum_{n} \,
   [ \, 1 - 2 f(\tau_n,\kperp,0) \, ] \,
   S(\tau_n,\kperp)
   \label{e:fsolII} \\
   & \qquad \times
   \Theta(\tau_n - \tau_0) \, \Theta(\tau - \tau_n) \>.
   \notag
\end{align}
Here $\tau_n$ are solutions of the equation
\begin{equation}\label{e:taun}
   \keta + e \, [ \, A(\tau) - A(\tau_n) \, ]
   =
   0 \>,
   \qquad \text{for $\tau_0 < \tau_n \le \tau$.}
\end{equation}
In order to step out $f(\tau,\kperp,\keta)$ as a function of $\tau$, we first solve \eqref{e:fsolII} at $\keta = 0$,
\begin{align}
   f(\tau,\kperp,0)
   &=
   \sum_{n} \,
   [ \, 1 - 2 f(\tau_n,\kperp,0) \, ] \,
   S(\tau_n,\kperp)
   \label{e:fsolzero} \\
   & \qquad \times
   \Theta(\tau_n - \tau_0) \, \Theta(\tau - \tau_n) \>,
   \notag
\end{align}
where now $\tau_n$ is a solution of the equation $A(\tau_n) = A(\tau)$, for $\tau_0 < \tau_n \le \tau$.  One such solution is for $\tau_n = \tau$.  Selecting out this case, and setting $\Theta(0)=1/2$, Eq.~\eqref{e:fsolzero} becomes
\begin{equation}\label{e:ftaujpkperpzero}
   f(\tau,\kperp,0)
   =
   \frac{
      S(\tau,\kperp)/2
      +
      \sum_{\tau_n < \tau} P(\tau_n,\kperp) \, S(\tau_n,\kperp) }
        { 1 + S(\tau,\kperp) } \>.
\end{equation}
We show a plot of $f(\tau,\kperp,0)$ in Fig.~\ref{f:0} for a typical case.  With $f(\tau,\kperp,0)$ now known, we can solve Eq.~\eqref{e:fsolII} for the full $f(\tau,\kperp,\keta)$.
%
%
%
%

%
%
\subsection{Maxwell's equations}
\label{ss:maxwell}

The only non-vanishing components of $F_{\mu\nu}$ are given in Eq.~\eqref{e:Fcontradef}, so Maxwell's equation \eqref{e:Maxwell} in boost invariant coordinates is given by
\begin{equation}\label{e:maxwellcovII}
   \partial_\tau E(\tau)
   =
   \tau J^{\eta}(\tau)
   =
   - J_{\eta}(\tau) / \tau \>,
\end{equation}
with $E(\tau) = - [ \partial_{\tau} A(\tau) ]/\tau$.
There are two types of currents, convection currents arising from the flow of particles and vacuum polarization currents,
\begin{equation}\label{e:JJconJpol}
   J_{\eta}(\tau)
   =
   J_{\eta}^{\text{con}}(\tau)
   +
   J_{\eta}^{\text{pol}}(\tau) \>.
\end{equation}
In a general frame, the convective current is given by the charge $e$ times the $\eta$-component of the particle current $N^{\mu}$ given in Eq.~\eqref{e:pcurrentdef},
\begin{equation}\label{e:Jcon}
   J_{\eta}^{\text{con}}(\tau)
   =
   e \, N_{\eta}(\tau)
   =
   e \int \rD k \, \keta \, f( \tau, \kperp, \keta ) \>.
\end{equation}
Inserting the result for $f( \tau, \kperp, \keta )$ from Eq.~\eqref{e:fsol} and integrating over $\keta$ gives
\begin{align}
   J_{\eta}^{\text{con}}(\tau)/\tau
   &=
   \frac{e R}{(2\pi)^2}
   \int_{0}^{\infty} \!\!\!\! \kperp \, \rd \kperp
   \int_{\tau_0}^{\tau} \!\! \rd \tau' \,
   \frac{ [ \, \pk_{\eta}(\tau',\tau)/\tau \, ]}
        { \omega_{\kperp}(\tau',\tau) } \,
   \Bigl ( \frac{\tau'}{\tau} \Bigr )
   \notag \\
   & \qquad \times
   P(\tau',\kperp) \,
   | E(\tau') | \,
   S(\tau',\kperp) \>,
   \label{e:JconIII}
\end{align}
where we have put
\begin{align}
   \pk_{\eta}(\tau',\tau)
   &=
   e \, [ \, A(\tau') - A(\tau) \, ] \>,
   \label{e:pkttomegaiidefs} \\
   \omega_{\kperp}(\tau',\tau)
   &=
   \sqrt{ \kperp^2 + [ \, \pk(\tau',\tau)/\tau \, ]^2 + M^2 } \>.
   \notag
\end{align}
The polarization current is determined by finding the appropriate current which, when added to the convection current, gives energy conservation.  In Section~\ref{ss:energymomentum} below, we found this current to be
\begin{align}
   &J_{\eta}^{\text{pol}}(\tau)/\tau
   =
   \text{sgn} [ \, E(\tau) \, ] \,
   \frac{ e R }{(2\pi)^2}
   \label{e:Jpol} \\ & \qquad \times
   \int_{0}^{\infty} \!\!\!\! \kperp \, \rd \kperp \,
   \omega_{\kperp,0}(\tau) \,
   P(\tau,\kperp) \,
   S(\tau,\kperp) \>,
   \notag
\end{align}
in agreement with Eq.~(5.8) in Cooper et.~al.~\cite{r:Cooper:1993uq}.

%
%
\subsection{Particle creation}
\label{ss:particlecreation}

The density of particles plus antiparticles at time $\tau$ is given by Eq.~\eqref{e:Ndefbi}.  Substituting our solution \eqref{e:fsol} into this equation gives
\begin{align}
   &\tau N_0(\tau)
   =
   \frac{R}{(2\pi)^2} \,
   \int_{0}^{\infty} \!\!\! \kperp \, \rd \kperp \!\!
   \int_{-\infty}^{+\infty} \!\!\! \rd \keta \,
   f(\tau,\kperp,\keta)
   \label{e:Ntau} \\
   &=
   \frac{e R}{(2\pi)^2}
   \int_{0}^{\infty} \!\!\!\!\! \kperp \, \rd \kperp \!\!
   \int_{\tau_0}^{\tau} \!\! \rd \tau' \, \tau'
   P(\tau',\kperp) \,
   | E(\tau') | \,
   S(\tau',\kperp) \>.
   \notag
\end{align}
The rate of production of particles plus antiparticles can be obtained by  differentiating \eqref{e:Ntau} with respect to $\tau$ and using the BV equation \eqref{e:BVeqII}.  This gives
\begin{equation}\label{e:Nrate}
   \frac{\rd [ \, \tau N_0(\tau) \, ]}{\tau \, \rd \tau}
   =
   \frac{ e R \, | E(\tau) | }{(2\pi)^2} \!\!
   \int_{0}^{\infty} \!\!\!\!\! \kperp \, \rd \kperp
   P(\tau,\kperp) \, S(\tau,\kperp) \>.
\end{equation}

\begin{figure}[t!]
   \centering
   \includegraphics[width=0.9\columnwidth]{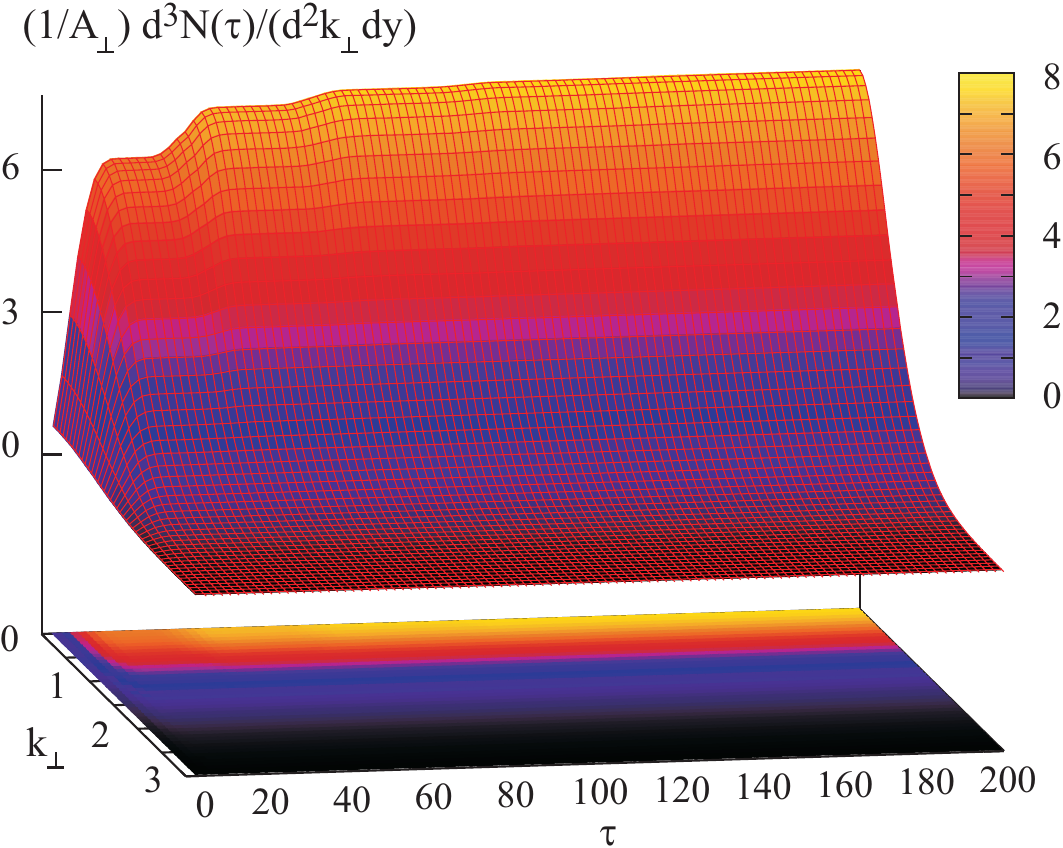}
   \caption{\label{f:7}(Color online)
   Time evolution of the transverse particle plus antiparticle distribution function,
   $\rd^3 N(\tau)/ \rd^2 k_{\perp} \rd y$,
   for boost-invariant coordinates in \threeplusone\ dimensions.
   See also Ref.~\onlinecite{ref:BVmovies}.}
\end{figure}

The particle production in terms of rapidity variables is obtained by substituting \eqref{e:fsol} into \eqref{e:dN5rapidII}.  This gives
\begin{align}\label{e:dN5rapidIII}
   &\frac{\rd^5 N}{\rd^2 x_{\perp} \rd^2 k_{\perp} \rd y}
   \\
   & \qquad =
   \frac{e R}{(2\pi)^3} \,
   \int_{\tau_0}^{\tau} \!\!\! \rd \tau' \,
   \tau' \,
   P(\tau',\kperp) \,
   | E(\tau') | \,
   S(\tau',\kperp) \>.
   \notag
\end{align}
A picture of this distribution is shown in Fig.~\ref{f:7} as a function of $\tau$.  Integrating \eqref{e:dN5rapidIII} over $\kperp$ gives
\begin{align}
   &\frac{1}{A_{\perp}}\frac{\rd N}{\rd y}
   \equiv
   \tau N_0(\tau)
   \label{e:dNdy} \\
   &=
   \frac{e R}{(2\pi)^2}
   \int_{0}^{\infty} \!\!\!\!\! \kperp \, \rd \kperp \!\!
   \int_{\tau_0}^{\tau} \!\! \rd \tau' \, \tau'
   P(\tau',\kperp) \,
   | E(\tau') | \,
   S(\tau',\kperp) \>,
   \notag
\end{align}
where $A_{\perp}$ is the perpendicular collision area.
%
%
\begin{figure}[t!]
   \centering
   \includegraphics[width=0.9\columnwidth]{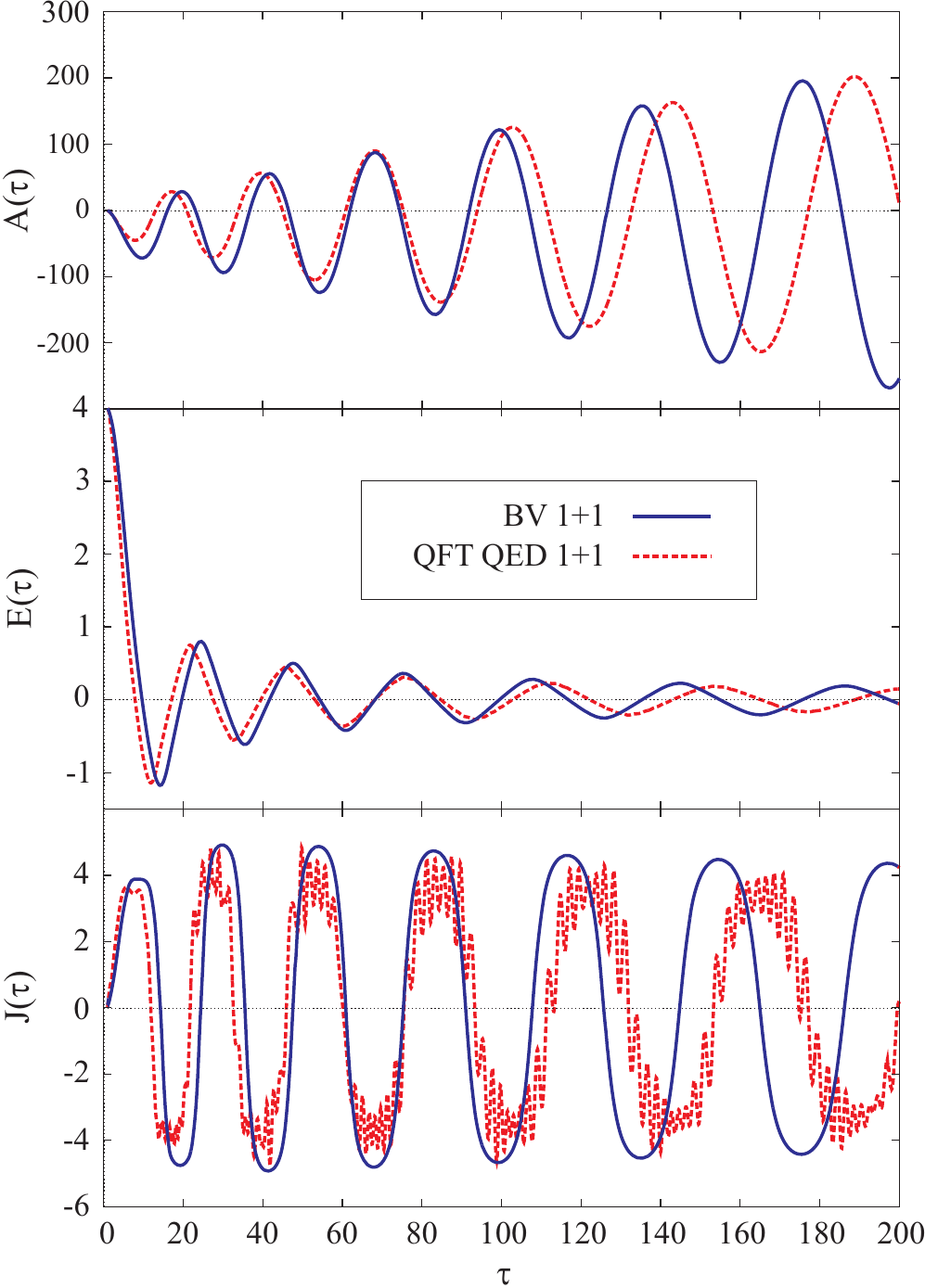}
   \caption{\label{f:1}(Color online)
   Proper-time evolution of the electromagnetic fields $A(\tau)$ and
   $E(\tau)$ and the electric current $J_{\eta}(\tau)$ for
   boost-invariant coordinates in \oneplusone dimensions.
   Solutions of the BV equation are compared with
   results from the quantum field theory (QFT) calculation discussed in
   Ref.~\onlinecite{r:Mihaila:2008dp}.
   Here we choose $M=1$, $e=1$, $A(\tau_0)=0$ and $E(\tau_0)=4$.}
\end{figure}
%
%
\begin{figure}[t!]
   \centering
   \includegraphics[width=0.9\columnwidth]{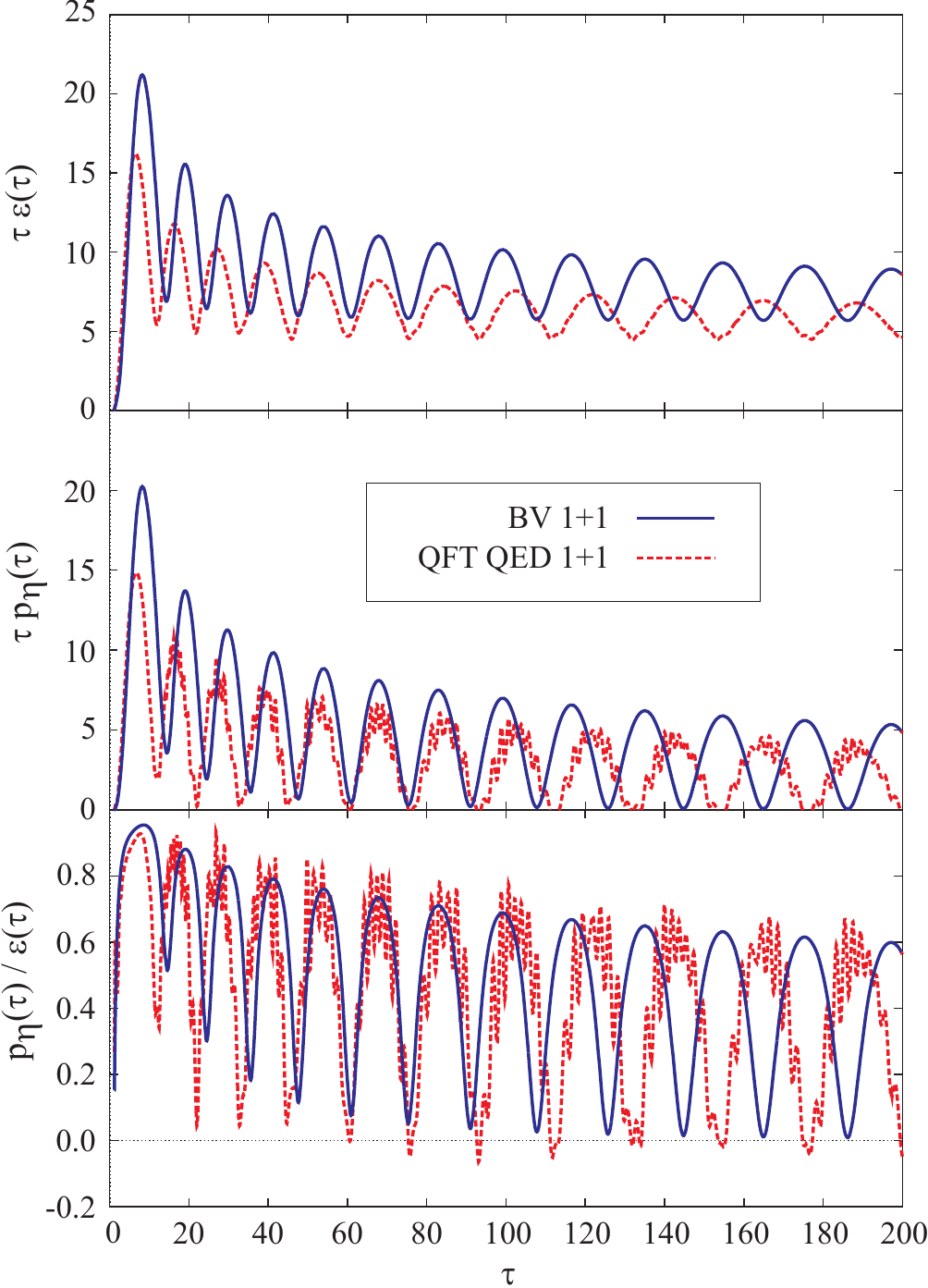}
   \caption{\label{f:2}(Color online)
   Proper-time evolution of the energy-momentum tensor (matter energy
   and longitudinal pressure) for boost-invariant coordinates in
   \oneplusone dimensions.  Parameters are the same as in Fig.~\ref{f:1}
   }
\end{figure}
%
%

%
%
\subsection{Energy-momentum tensor}
\label{ss:energymomentum}

The field energy-momentum tensor density is given by \eqref{e:enrgmomfield}.  For our case in boost-invariant coordinates, it is diagonal and given by
\begin{equation}\label{e:tfield}
   \Theta_{\mu\nu}
   =
   \frac{1}{2} \, \Diag{ E^2,E^2,\rho^2 E^2,-\tau^2E^2 } \>.
\end{equation}
The matter energy-momentum tensor is given by \eqref{e:penergymom} which we write here as
\begin{align}
   t_{\mu\nu}(\tau)
   &=
   \int \rD k \>
   \pk_{\mu} \, \pk_{\nu} \, f(\tau,\kperp,\keta)
   \label{e:tmat} \\
   &\defby
   \Diag{
      \epsilon, p_{\rho},
      \rho^2 \, p_{\theta},
      \tau^2 p_{\eta} } \>.
\end{align}
So in the boost-invariant system, the matter energy and pressures are given by
\begin{subequations}\label{e:epppdefs}
\begin{align}
   \epsilon
   &=
   \int \rD k \>
   \omega_{\kperp,\keta}^2 \,
   f( \tau, \kperp, \keta ) \>,
   \label{e:energydef} \\
   p_{\rho}
   &=
   \int \rD k \> \pk_{\rho}^2 \,
   f( \tau, \kperp, \keta ) \>,
   \label{e:Prhodef} \\
   p_{\theta}
   &=
   \int \rD k \>
   ( \pk_{\theta} / \rho)^2 \,
   f( \tau, \kperp, \keta ) \>,
   \label{e:Pthetadef} \\
   p_{\shortparallel}
   &=
   \int \rD k \>
   (\keta/\tau)^2 \,
   f( \tau, \kperp, \keta ) \>.
   \label{e:Petadef}
\end{align}
\end{subequations}
Inserting the result for $f( \tau, \kperp, \keta )$ from Eq.~\eqref{e:fsol} and integrating over $\keta$ gives, for \eqref{e:energydef} and \eqref{e:Petadef},
\begin{subequations}\label{e:energyPetaresults}
\begin{align}
   \epsilon
   &=
   \frac{e R}{(2\pi)^2}
   \int_{0}^{+\infty} \!\!\!\!\!\! \kperp \, \rd \kperp
   \int_{\tau_0}^{\tau} \!\! \rd \tau' \,
   \omega_{\kperp}(\tau',\tau) \,
   \Bigl ( \frac{\tau'}{\tau } \Bigr )
   \notag \\
   & \qquad \times
   P(\tau',\kperp) \,
   | E(\tau') | \,
   S(\tau',\kperp) \>,
   \label{e:energyresult} \\
   p_{\shortparallel}
   &=
   \frac{e R}{(2\pi)^2}
   \int_{0}^{+\infty} \!\!\!\!\!\! \kperp \, \rd \kperp
   \int_{\tau_0}^{\tau} \!\! \rd \tau' \,
   \frac{ [ \, \pk_{\eta}(\tau',\tau)/\tau \, ]^2 }
        { \omega_{\kperp}(\tau',\tau) }
   \Bigl ( \frac{\tau'}{\tau } \Bigr )
   \notag \\
   & \qquad \times
   P(\tau',\kperp) \,
   | E(\tau') | \,
   S(\tau',\kperp) \>.
   \label{e:Petaresult}
\end{align}
\end{subequations}
where $\pk_{\eta}(\tau',\tau)$ and $\omega_{\kperp}(\tau',\tau)$ are given in Eqs.~\eqref{e:pkttomegaiidefs}.

%
%
\begin{figure}[t!]
   \centering
   \includegraphics[width=\columnwidth]{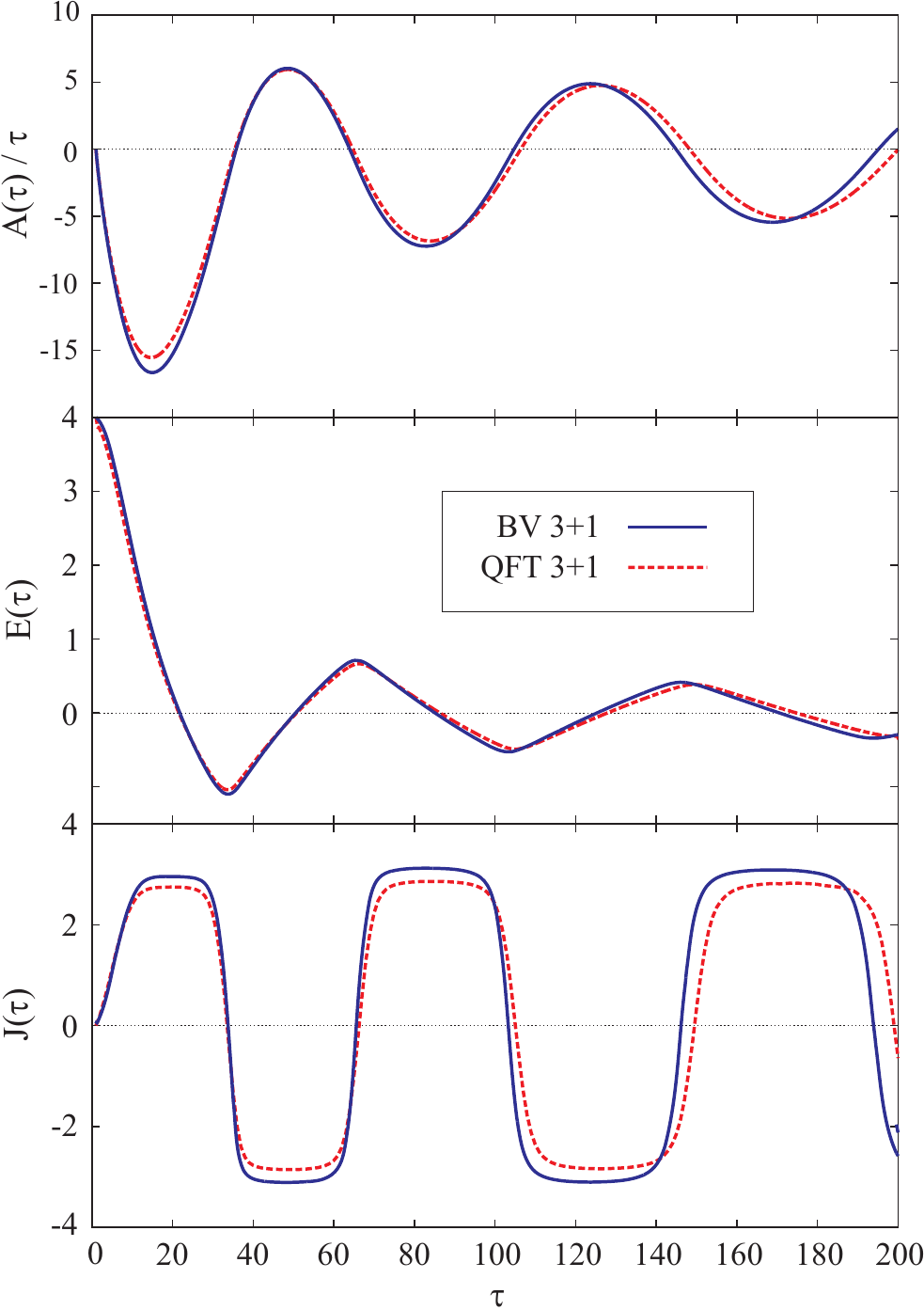}
   \caption{\label{f:3}(Color online)
   Time evolution of the electromagnetic fields $A(\tau)/\tau$ and
   $E(\tau)$ and the electric current $J_{\eta}(\tau)$ for
   boost-invariant coordinates in \threeplusone dimensions.
   Solutions of the BV equation are compared with
   results from the quantum field theory (QFT) calculation discussed in
   Ref.~\onlinecite{r:Mihaila:2009tg}.
   Here we choose $M=1$, $e=1$, $A(\tau_0)=0$ and $E(\tau_0)=4$.}
\end{figure}
%
%
\begin{figure}[t!]
   \centering
   \includegraphics[width=\columnwidth]{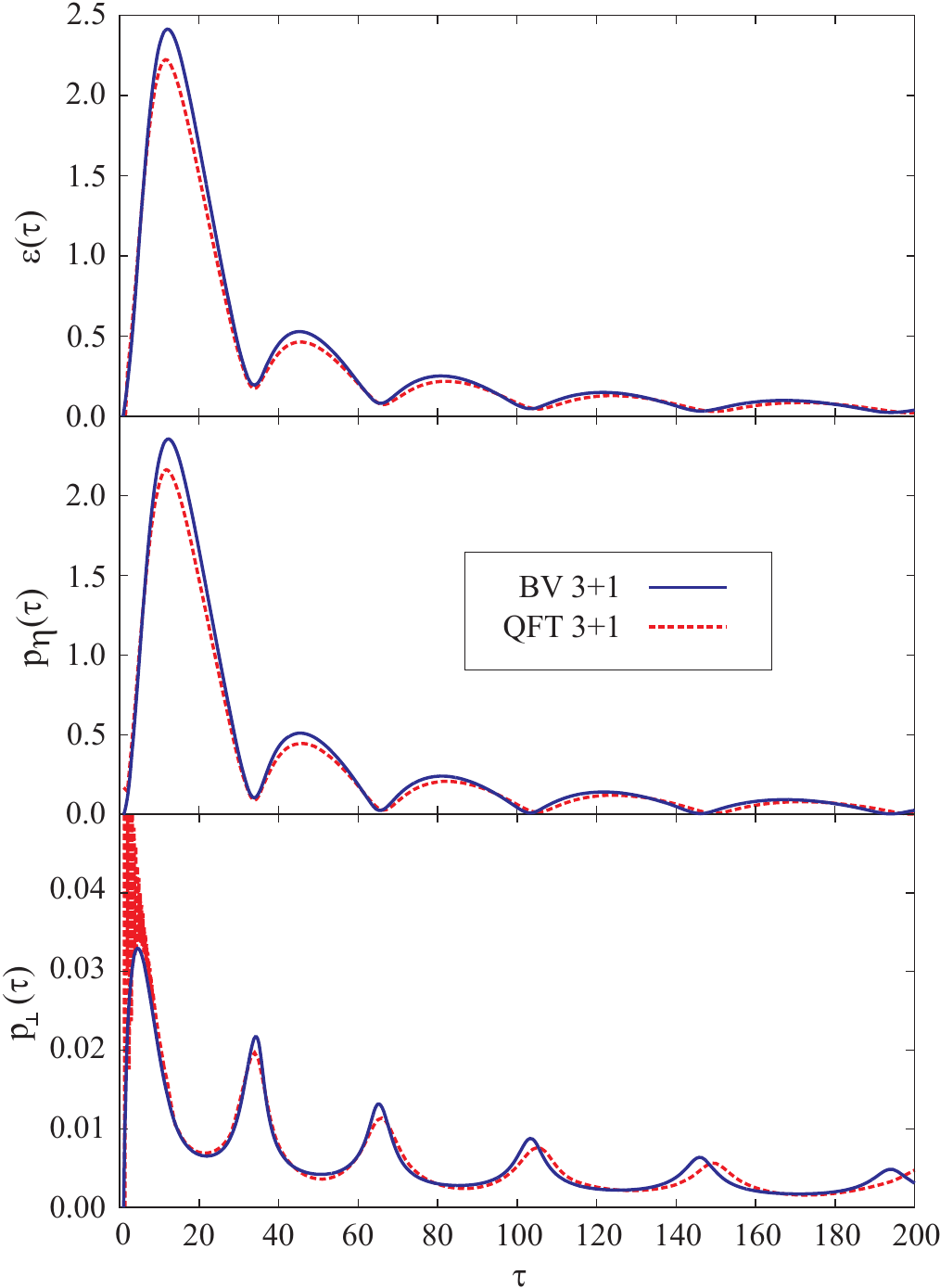}
   \caption{\label{f:4}(Color online)
   Time evolution of the matter energy ($\epsilon$), longitudinal pressure
   ($p_{\shortparallel}$), and transverse pressure ($p_{\perp}$)
   for boost-invariant coordinates in \threeplusone dimensions, with the same
   parameters as in Fig.~\ref{f:3}.}
\end{figure}
%
%

Multiplying the BV equation \eqref{e:BVeqII} by $\omega_{\kperp,\keta}^2$, and integrating over $\rD k$, gives
\begin{align}
   &\frac{R}{(2\pi)^2}
   \int_{0}^{\infty} \!\!\! \kperp \rd \kperp \!\!
   \int_{-\infty}^{+\infty} \!\!\! \rd \keta \,
   \biggl \{ \,
      \frac{\omega_{\kperp,\keta}(\tau)}{\tau} \,
      \frac{\partial f(\tau,\kperp,\keta)}{\partial \tau}
      \notag \\ & \qquad
      +
      e E(\tau) \,
      \omega_{\kperp,\keta}(\tau) \,
      \frac{\partial f(\tau,\kperp,\keta)}{\partial \keta} \,
   \biggr \}
   \label{e:BVeqIIint}  \\
   &=
   \frac{R \, | e E(\tau) |}{(2\pi)^2} \,
   \int_{0}^{\infty} \!\!\!\! \kperp \, \rd \kperp \,
   \omega_{\kperp,0}(\tau) \,
   P(\tau,\kperp) \,
   S(\tau,\kperp) \>.
   \notag
\end{align}
For the first term in \eqref{e:BVeqIIint}, we integrate by parts and note that
\begin{equation}\label{e:deromegatau}
\begin{split}
   \tau \,
   \frac{\partial}{\partial \tau} \,
   \biggl (
      \frac{\omega_{\kperp,\keta}(\tau)}{\tau}
   \biggr )
   &=
   -
   \frac{\omega_{\kperp,\keta}(\tau)}{\tau}
   +
   \frac{\partial\omega_{\kperp,\keta}(\tau)}{\partial \tau}
   \\
   &=
   -
   \frac{\omega_{\kperp,\keta}(\tau)}{\tau}
   -
   \frac{ (\keta/\tau)^2 }{ \tau \, \omega_{\kperp,\keta}(\tau) } \>.
\end{split}
\end{equation}
So the first term becomes simply
\begin{equation}\label{e:firstterm}
   \frac{\partial \epsilon}{\partial \tau}
   +
   \frac{\epsilon + p_{\shortparallel}}
        {\tau} \>.
\end{equation}
For the second term in \eqref{e:BVeqIIint}, we integrate by parts over $\keta$ and get
\begin{equation}\label{e:secondterm}
   - e \, \frac{E(\tau)}{\tau} \,
   \int \rD k \, \keta \,
   f(\tau,\kperp,\keta)
   =
   - \frac{E(\tau) \, J_{\eta}^{\text{con}}(\tau)}{\tau} \>,
\end{equation}
where the convective current is given by \eqref{e:Jcon}.
The last term in Eq.~\eqref{e:BVeqIIint} can be written as
\begin{equation}\label{e:lastterm}
   E(\tau) \, J_{\eta}^{\text{pol}}(\tau) / \tau \>,
\end{equation}
where the polarization current $J_{\eta}^{\text{pol}}(\tau)$ is given by
\begin{align}
   &J_{\eta}^{\text{pol}}(\tau)/\tau
   =
   \text{sgn} [ \, E(\tau) \, ] \,
   \frac{ e R }{(2\pi)^2}
   \label{e:JpolII} \\ & \qquad \times
   \int_{0}^{\infty} \!\!\!\! \kperp \, \rd \kperp \,
   \omega_{\kperp,0}(\tau) \,
   P(\tau,\kperp) \,
   S(\tau,\kperp) \>,
   \notag
\end{align}
which is what we quoted in Eq.~\eqref{e:Jpol} in Section~\ref{ss:maxwell}.
Combining the results in Eqs.~\eqref{e:firstterm}, \eqref{e:secondterm}, and \eqref{e:lastterm}, and noting that the total current is given by $J_{\eta}(\tau) = J_{\eta}^{\text{con}}(\tau) + J_{\eta}^{\text{pol}}(\tau)$, the energy density and longitudinal pressure conservation equation reads
\begin{equation}\label{e:eostateI}
   \frac{ \partial \epsilon }{\partial \tau}
   +
   \frac{\epsilon + p_{\shortparallel} }{\tau}
   =
   \frac{E(\tau) \, J_{\eta}(\tau)}{\tau}
   =
   -
   \frac{ \partial [ \, E^2 / 2 \, ]}{\partial \tau}
\end{equation}
where we have used Maxwell's equation \eqref{e:maxwellcovII}.  The total energy density and longitudinal pressure is given by
\begin{equation}\label{e:Ttotals}
   \calE
   =
   \epsilon
   +
   E^2 / 2\>,
   \qquad
   \calP_{\shortparallel}
   =
   p_{\shortparallel}
   -
   E^2 / 2 \>.
\end{equation}
Multiplying Eq.~\eqref{e:eostateI} by $\tau$ gives an equation of state:
\begin{equation}\label{e:eostateII}
   \partial_\tau ( \, \tau \calE \, )
   +
   \calP_{\shortparallel}
   =
   0 \>.
\end{equation}

The $p_{\rho}(\tau)$ and $p_{\theta}(\tau)$ pressures are equal.  We can prove this by introducing the integration variables $\bar{\pk}_{\theta} = \pk_{\theta}/\rho$ and $\bar{\pk}_{\eta} = \pk_{\eta} / \tau$, and putting
\begin{align}
   \int \rD k \>
   &=
   R
   \int_{0}^{+\infty}
   \frac{ \kperp \, \rd \kperp }
        { 2\pi }
   \int_{-\infty}^{+\infty} \frac{ \rd \keta }{ 2\pi } \,
   \frac{1}{ \tau \, \omega_{\kperp,\keta} }
   \label{e:notice} \\
   &=
   R
   \iiint_{-\infty}^{+\infty} \!
   \frac{ \rd \pk_{\rho} \, \rd \pk_{\theta} \, \rd \keta }
        { (2\pi)^3 } \,
   \frac{1}{\tau \rho \,
      \omega_{\pk_{\rho},\bar{\pk}_{\theta},\keta} }
   \notag \\
   &=
   R
   \iiint_{-\infty}^{+\infty} \!
   \frac{ \rd \pk_{\rho} \, \rd \bar{\pk}_{\theta} \, \rd \bar{\pk}_{\eta} }
        { (2\pi)^3 } \,
   \frac{1}{
      \omega_{\pk_{\rho},\bar{\pk}_{\theta},\bar{\pk}_{\eta}} } \>,
   \notag
\end{align}
where now
\begin{equation}\label{e:omegakkk}
   \omega_{ \pk_{\rho}, \bar{\pk}_{\theta}, \bar{\pk}_{\eta} }
   =
   \sqrt{ \pk_{\rho}^2 + \bar{\pk}_{\theta}^2
          + \bar{\pk}_{\eta}^2 + M^2 } \>.
\end{equation}
From \eqref{e:Prhodef} and \eqref{e:Pthetadef}, we find for the pressures
\begin{align*}
   p_{\rho}(\tau)
   &=
   R
   \iiint_{-\infty}^{+\infty} \!
   \frac{ \rd \pk_{\rho} \, \rd \bar{\pk}_{\theta} \,
          \rd \bar{\pk}_{\eta} }
        { (2\pi)^3 } \,
   \frac{\pk_{\rho}^2 \,
         f( \tau, \pk_{\rho},\bar{\pk}_{\theta},\bar{\pk}_{\eta} )}{
      \omega_{\pk_{\rho},\bar{\pk}_{\theta},\bar{\pk}_{\eta}} } \>,
   \\
   p_{\theta}(\tau)
   &=
   R
   \iiint_{-\infty}^{+\infty} \!
   \frac{ \rd \pk_{\rho} \, \rd \bar{\pk}_{\theta} \,
          \rd \bar{\pk}_{\eta} }
        { (2\pi)^3 } \,
   \frac{\bar{\pk}_{\theta}^2 \,
         f( \tau, \pk_{\rho},\bar{\pk}_{\theta},\bar{\pk}_{\eta} )}{
      \omega_{\pk_{\rho},\bar{\pk}_{\theta},\bar{\pk}_{\eta}} } \>,
\end{align*}
so $p_{\rho}(\tau) = p_{\theta}(\tau)$, as we claimed.  Including the field pressure, we see that the total pressures also satisfy the relation $\calP_{\rho}(\tau) = \calP_{\theta}(\tau)$, as required by conservation of the energy-pressure tensor.

For the transverse pressure, we have
\begin{align}
   p_{\perp}(\tau)
   &=
   p_{\rho}(\tau)
   +
   p_{\theta}(\tau)
   =
   \int \rD k \> \kperp^2 \,
   f( \tau, \kperp, \keta )
   \label{e:Pperp} \\
   &=
   \frac{e R}{(2\pi)^2}
   \int_{0}^{+\infty} \!\!\!\!\!\! \kperp \, \rd \kperp
   \int_{\tau_0}^{\tau} \!\! \rd \tau' \,
   \frac{ \kperp^2 }
        { \omega_{\kperp}(\tau',\tau) }
   \Bigl ( \frac{\tau'}{\tau } \Bigr )
   \notag \\
   & \qquad \times
   P(\tau',\kperp) \,
   | E(\tau') | \,
   S(\tau',\kperp) \>.
   \notag
\end{align}
The shear pressure vanishes.  In the next section, we compare results of solving the BV equation with a quantum field theory calculation in both \oneplusone\ and \threeplusone\ dimensions.

%
%
\begin{figure}[t!]
   \centering
   \includegraphics[width=0.9\columnwidth]{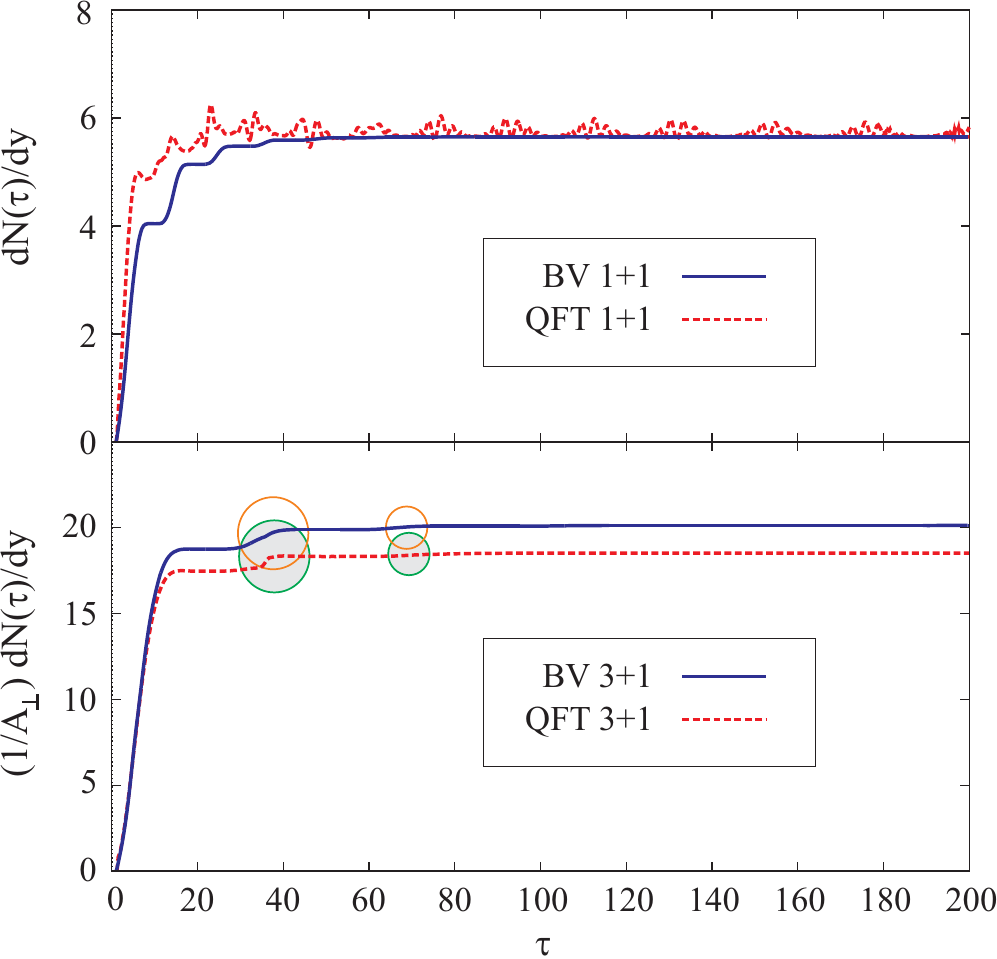}
   \caption{\label{f:5}(Color online)
   Time evolution of the particle plus antiparticle density
   distribution function for boost-invariant coordinates
   in \oneplusone\ dimensions, $\rd N(\tau)/\rd y $, and
   in \threeplusone\ dimensions, $( 1/ A_{\perp} ) \, \rd N(\tau)/\rd y $,
   respectively.
   In \threeplusone\ dimensions,
   the BV calculation predicts slightly more particle production.}
\end{figure}
%
%
\begin{figure}[t!]
   \centering
   \includegraphics[width=0.9\columnwidth]{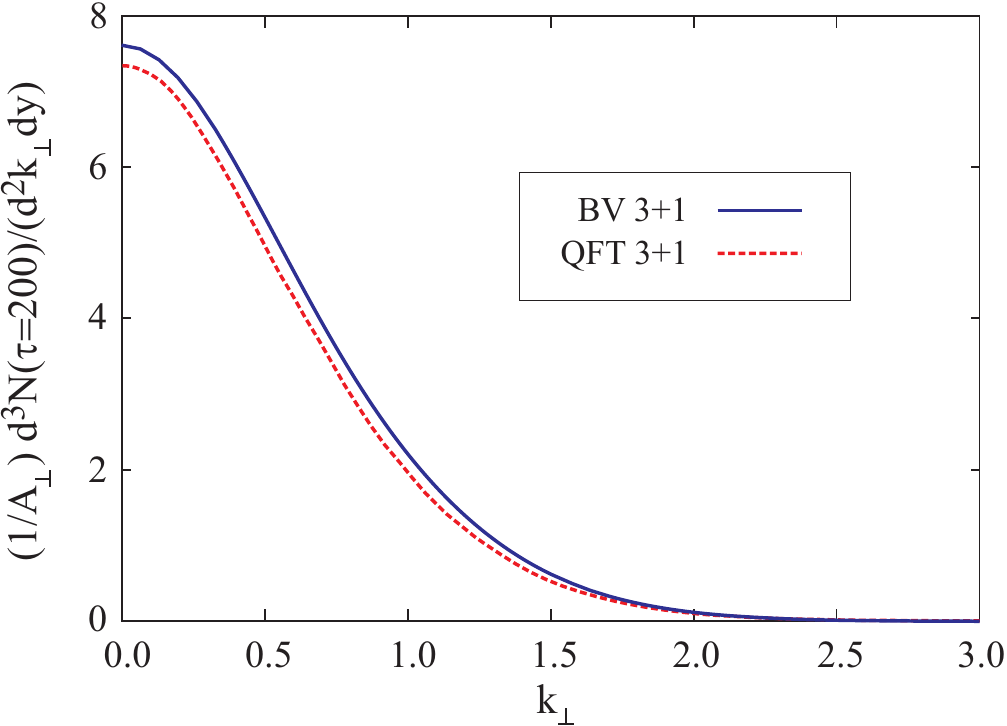}
   \caption{\label{f:6}(Color online)
   Long proper-time ($\tau=200$) transverse particle plus antiparticle distribution function,
   $( 1/ A_{\perp} ) \, \rd^3 N(\tau=200) / \rd^2 k_{\perp} \rd y$,
   for boost-invariant coordinates in \threeplusone\ dimensions.}
\end{figure}
%
%
%
%

%
%
\section{Numerical methods and results}
\label{s:numerical}

The numerical procedure is as follows: we set $R=2$ and $R=4$ for \oneplusone\ and \threeplusone\ dimensions, respectively, and choose units such that $\hbar = 1$. We set $M=1$ and $e=1$, and take $\tau_0 = 1/M = 1$.  Then we set $A(\tau_0) = 0$, and choose a value of $E_0$.  We set up a grid of values of $\kperp$ between $0$ and $\pk_{\kperp \text{max}}$, and compute initial values for $J_{\eta}(\tau_0)$ and $f(\tau_0,\kperp,0)$.  We also compute a value for $\rd J_{\eta} / \rd \tau$ at $\tau_0$.  We can then take a fourth-order Runge-Kutta step in $\tau$ to find new values for $A(\tau)$ and $E(\tau)$, using a linear interpolator for values of $J_{\eta}(\tau)$.  We then compute values for $f(\tau,\kperp,0)$ from Eq.~\eqref{e:fsolzero}, $J_{\eta}(\tau)$ using Eq.~\eqref{e:JconIII}, and $\rd J_{\eta} / \rd \tau$ at the new value of $\tau$, and proceed in this way until we reach the final value of $\tau$.  This method does not require computation of the full function $f(\tau,\kperp,\keta)$ at the expense of an additional integral over $\tau$; however $f(\tau,\kperp,\keta)$ can be computed at any point along the way.

We consider the case when $E_0=4$, and compare the Boltzman-Vlasov (BV) results with two sets of recent quantum
field theory (QFT) calculations done by us in \oneplusone\ and \threeplusone\ dimensional QED~\cite{r:Mihaila:2008dp,r:Mihaila:2009tg}.  Comparisons for \oneplusone-dimensional QED are shown in Fig.~\ref{f:1} for $A(\tau)$, $E(\tau)$, and $J(\tau)$, and in Fig.~\ref{f:2} for components of the energy-momentum tensor.  The BV calculation
misses the fine structure noticed in the oscillations of the QFT electric current calculation, which has some features of quantum tunneling in a two-well potential (see Ref.~\onlinecite{r:Bender:1985fk}), but otherwise is close in magnitude.  The two calculations get out of phase for large times, but this does not affect the calculation of the particle production which is dominated by the early-time dynamics. The BV calculation predicts larger energy density and longitudinal pressure, but about the same ratios of energy density to pressure.

In Figs.~\ref{f:3} and~\ref{f:4} we compare the BV results to the QFT results for \threeplusone-dimensional QED.  We note that the fields, currents, energy density, and pressures all track very well together. The agreement between the BV and QFT calculations is better in \threeplusone\ than in \oneplusone\ dimensions, suggesting that the extra degrees of freedom perform some smoothing. In \threeplusone\ dimensions we do not observe dephasing between the BV and QFT results at late times, at least as far as our calculations were carried out. Again, the BV calculation predicts larger values for the energy density and longitudinal pressure, but the transverse pressure for both calculations are fairly close to each other.  There is no fine structure present in the \threeplusone\ QFT results for the electric current, as discussed in Ref.~\onlinecite{r:Mihaila:2009tg}.

In Fig.~\ref{f:5}, we show the particle plus antiparticle production per unit rapidity for the two calculations. In \oneplusone\ dimensions, the particle plus antiparticle production per unit rapidity for both calculations are very close, aside from the fine structure. In \threeplusone\ dimensions, the BV calculation predicts a slightly larger production than in the QFT results, which is consistent with the fact that the BV electric current depicted in Fig.~\ref{f:3} is slightly larger than the QFT current. Just, as in QFT, particles are being created corresponding to the field gradients, with the major contribution coming from the initial field gradient.  Subsequent smaller step increases are observed before the particle density saturates.

Comparison of the late time ($\tau=200$) transverse particle plus antiparticle distributions for the BV and QFT calculations are shown in Fig.~\ref{f:6}.  The results are very close.  Finally, in Fig.~\ref{f:7}, we show the BV calculation for the entire time evolution of the transverse particle plus antiparticle distribution, which very similar to the one reported in Ref.~\onlinecite{r:Mihaila:2009tg}, except for an approximate 5-10\% difference in magnitude. For $\tau$ greater than about 80, there is no appreciable change in the shape of the distribution function, as expected, since by that time all particles have been produced by the field.

%
%
\section{Conclusions}
\label{s:conclusions}

We have presented here results of a non-equilibrium BV calculation of the time evolution of the quasiparticle distribution function for quarks in the presence of a proper-time evolving electric field with a Schwinger pair creation term in boost-invariant coordinates in \oneplusone\ and \threeplusone\ dimensions.  We have then compared these results with recent QFT calculations.  Our one-dimensional results agree with previous results in Ref.~\onlinecite{r:Cooper:1993uq} and give reasonable agreement with the field theory calculations when short time scales are averaged over.  What is initially surprising is that in \threeplusone\ dimensions, the short time scale fluctuations are not present in the field theory calculations so that agreement between the exact and the BV approximation for many macroscopic variables such as the time evolution of the electric field and the effective energy density and pressures are quite good.
%
%
The two methods differ in the particle production rate by about 5-10\%, mostly at low momentum transfers.

It is at first quite surprising that the BV results are so close to the QFT results. A first-principles approach to deriving a BV-like equation for the exact field theory equations in scalar electrodynamics in \oneplusone\ dimension has been given in Ref.~\onlinecite{r:Kluger:1998mi} where obtaining a local Vlasov source term from the non-local equation for the adiabatic number operator seemed to follow from phase decoherence of the quantum density matrix.  In \threeplusone\ dimensions we would imagine that this phase decoherence would occur more quickly than in \oneplusone\ dimensions, which would make the quantum to classical transition quite rapid.  This would then  be the reason why the semiclassical approach presented here works better in \threeplusone\ than in \oneplusone\ dimensions.

The fact that the BV calculations are computationally much faster than solving the field theory equations  makes them a good candidate for extending this work to the case of QCD, where the computer time required for a full QFT calculation can become prohibitive for an exhaustive investigation of the two SU(3) Casimir invariants parameter space.  If the BV approach with the  correct Schwinger source term proves to be as accurate in QCD as in QED then it would be very helpful in exploring parameter space so that the Casimirs dependence for the transverse distribution function can be better understood for the case when back-reaction is included.  We intend to explore this possibility in a subsequent publication.

%
%
\begin{acknowledgments}
This work was performed in part under the auspices of the United States Department of Energy. The authors would like to thank the Santa Fe Institute for its hospitality during the completion of this work.
\end{acknowledgments}
\vfill
%
%
\bibliography{johns}
%
%
\end{document}